\documentclass[12pt]{article}
\usepackage{amsmath}
\usepackage[dvips]{graphicx}
\begin{document}
\def\scri{
\unitlength=1.00mm
\thinlines
\begin{picture}(3.5,2.5)(3,3.8)
\put(4.9,5.12){\makebox(0,0)[cc]{$\cal J$}}
\end{picture}}

\def \th {\theta}
\def \ph {\varphi}
\def \ga {\gamma}
\def \sh {\sinh}
\def \ch {\cosh}
\def \s {\sin}
\def \c {\cos}
\def \ni {\noindent}

\begin{titlepage}

\noindent Stockholm\\
SUITP 00-17\\
December 2000

\vspace{2cm}

\begin{center}

{\Large 2+1 GRAVITY, CHAOS AND TIME MACHINES}

\vspace{12mm}

{\large Ingemar Bengtsson}\footnote{Email address: ingemar@physto.se. 
Supported by NFR.}

\vspace{6mm}

{\large Johan Br\"annlund}\footnote{Email address: jbr@physto.se}

\vspace{8mm}

{\sl Fysikum\\
Stockholm University\\
Box 6730, S-113 85 Stockholm, Sweden}

\vspace{10mm}

{\bf Abstract}

\end{center}

\vspace{6mm}

\noindent 2+1 gravity for spacetimes with topology ${\bf R} \times {\bf T}^2$ 
has been much studied. We add a description of how to extend these spacetimes 
across a Cauchy horizon into a region where the torus becomes Lorentzian. The 
result is a one parameter family of tori given by a geodesic in the 
"Teichm\"uller space" of Lorentzian tori. We describe this in detail. We also 
point out that if the modular group is regarded as part of the gauge group 
then these spacetimes offer a nice toy model for the dynamics of Bianchi IX models; 
in the region where the tori are spacelike the dynamics is described exactly 
by a hyperbolic billiard. On the other hand the modular group acts ergodically 
on the Teichm\"uller space of Lorentzian tori.

\end{titlepage}

\noindent {\bf 1. INTRODUCTION.}

\vspace{5mm}

\noindent The subject of 2+1 dimensional gravity looks {\it a priori} 
unpromising since---in the absence of matter---all spacetimes have 
constant curvature. Nevertheless it has been the subject of many investigations 
over the past twenty years or so. Indeed it is now widely recognized 
that it provides (when handled with taste!) surprisingly illuminating 
toy models of general relativity. Most of these investigations center 
on quantum gravity \cite{Carlip}, often from a Hamiltonian point of view, 
and as a result the spacetime properties of the models are receiving somewhat 
less attention than we think that they deserve. Here we intend to present 
some properties of 2+1 spacetimes with topology ${\bf R}\times {\bf T}^2$, 
regarded as quotients of Minkowski space. This perspective enables us 
to discuss what goes on in that region of spacetime where the torus becomes 
Lorentzian and closed timelike curves appear; existing treatments 
\cite{Moncrief} 
\cite{Hosoya} typically use the Hamiltonian ADM formalism and therefore do 
not go across the Cauchy horizon that bounds this region. The motivation 
for doing this is partly just curiosity, but partly a feeling that there is 
structure there which may well illuminate some features occurring in 3+1 
dimensions too---even if it will manifest itself in a different 
way in the latter case. Be that as it may a nice picture emerges; we can 
regard the entire spacetime as a geodesic in the Teichm\"uller space of 
tori. This space is the familiar upper half plane in the Riemannian case, 
and it is 1+1 dimensional de Sitter space in the Lorentzian case. 

The second point that we wish to bring up is that if the modular group 
is regarded 
as part of the gauge group then these spacetimes offer a nice toy model for 
the chaotic behaviour of Bianchi cosmologies. The dynamics of the latter 
has attracted attention for quite some time and many of its aspects are 
by now well understood. (There are many references, old \cite{Belinski} 
\cite{Charles}, new \cite{Hobill} and very new 
\cite{Ringstrom}.) In particular it is well known that the behaviour of 
Bianchi IX models close to the singularity can be approximated by a hyperbolic 
billiard, which is an archetypical chaotic system. In the literature the 
situation is often described by saying that chaotic behaviour appears when 
curvature becomes strong, although the precise meaning of the word "chaotic" 
here is a subject of some controversy. It is therefore of some interest 
that this kind of chaotic behaviour appears in 2+1 gravity with zero 
curvature, as a kind of global effect. A simplifying feature is that in 
our case the hyperbolic billiard captures the dynamics exactly. 

As additional motivation we note that both the points we raise are 
important for quantization. They also appear to be of interest in string 
theory---see ref. \cite{Moore}, but beware of some misunderstandings in 
that reference. 

The organization of the paper is as follows: In section 2 we construct our 
spacetimes by taking quotients of a region of 2+1 dimensional Minkowski space 
with the appropriate discrete isometry groups. This construction is well 
known \cite{Mess} \cite{Louko}. In section 3 we describe these spacetimes as 
a geodesic in a Teichm\"uller space; this is a new result as far as the 
region with closed timelike curves is concerned. Since the 
Teichm\"uller space of Lorentzian tori has caused some puzzlement in 
the past \cite{Yurtsever} we describe it in detail. In section 4 we describe 
the dynamics which results when taking the quotient of Teichm\"uller space 
with the modular group, and stress the analogy to mixmaster cosmology. 
We focus on the spectrum of closed geodesics since they are the skeleton 
on which chaos is built; actually a closed geodesic corresponds to a 
self-similar rather than a periodic spacetime. Our account is intended to 
be pedagogical (and to be helpful in section 5); all the hard results are 
well known to mathematicians \cite{Artin} \cite{Series} and to workers in 
quantum chaos \cite{Bogomolny}. In section 5 we discuss the 
action of the modular group on the Teichm\"uller space of Lorentzian tori. 
We show that it is ergodic. (In a general setting involving discrete groups 
acting on coset spaces formed from non-compact groups such phenomena are 
known to mathematicians, but our pedestrian treatment is original as far 
as we know.) In section 6 we sketch how our method works for locally de Sitter 
spacetimes \cite{Soda}, and comment on the higher genus case. Our 
conclusions are in section 7. 

\vspace{1cm}

\noindent {\bf 2. OUR SPACETIMES.}

\vspace{5mm}

\noindent Let ${\bf M}$ be a region of 2+1 dimensional Minkowski space and 
${\Gamma}$ a free discrete isometry group acting in a properly discontinuous 
way on this region. We want to choose ${\Gamma}$ so that the quotient space 
${\bf M}/{\Gamma}$ has the topology of a torus cross the real line. For a 
simply connected $M$ the quotient space has ${\Gamma}$ as its 
fundamental group. Therefore ${\Gamma}$ must be a free discrete group with 
two commuting generators. We also insist that the quotient space should 
contain a complete spacelike surface that is not crossed by any closed 
timelike (or null) curve. The solution to this problem is described, e.g., 
by Louko and Marolf \cite{Louko}. As generators of the discrete group we 
choose $g_1 = e^{{\xi}_1}$ and $g_2 = e^{{\xi}_2}$, that is to say 
exponentials of the two linearly independent commuting Killing vectors 

\begin{equation} {\xi}_1 = {\alpha}J_{xt} + {\beta}P_y \hspace{8mm} 
{\xi}_2 = {\gamma}J_{xt} + {\delta}P_y \ ; \hspace{8mm} {\alpha}{\delta} - 
{\beta}{\gamma} > 0. \end{equation}

\noindent Here $J_{xt}$ is a Lorentz boost, $P_y$ is a translation and 
${\alpha}, 
{\beta}, {\gamma}$ and ${\delta}$ are real numbers. This is the most general 
solution, except for the obvious static case that we ignore. The group 
${\Gamma}$ will contain all group elements of the form $e^{\xi}$, where 

\begin{equation} {\xi} = (n_1{\alpha} - n_2{\gamma})J_{xt} + (n_1{\beta} - 
n_2{\delta})P_y \ , \hspace{6mm} n_1, n_2 \in {\bf Z} \ . \label{xi} 
\end{equation}

\noindent Here $n_1$ and $n_2$ are arbitrary integers. We observe that 
${\Gamma}$ 
will contain pure boosts if and only if ${\beta}/{\delta}$ is rational, 
and pure translations if and only if ${\alpha}/{\gamma}$ is rational. Note 
also that in any case the action of ${\Gamma}$ on the line $x = t = 0$ is 
problematic; if a pure boost is present it has a line of fixed points there, 
and if not the action of ${\Gamma}$ on this line is ergodic. Hence we see 
why the covering space $M$ is taken to be a subset of 2+1 dimensional 
Minkowski space only.

Since the Killing vectors ${\xi}_1$ and ${\xi}_2$ commute they form surfaces, 
namely 

\begin{equation} t^2 - x^2 = {\tau}^2 \equiv - {\sigma}^2 \ , \end{equation}

\noindent where ${\tau}^2$ is some constant (not necessarily positive; if 
it is not then ${\sigma}^2$ is positive). These surfaces are left 
invariant by the group 
${\Gamma}$, they foliate Minkowski space, they are intrinsically flat 
and their mean curvature is constant. They turn into tori when we take the 
quotient with ${\Gamma}$. From now on we take $M$ to be the union of regions 
I and II of Minkowski space, as defined in figure 1. This means that our 
quotient spaces will be geodesically incomplete. If we did not restrict 
$M$ in this way we would obtain what Louko and Marolf \cite{Louko} accurately 
describe as a "modest generalization of Misner space"; as far as we can see 
there is nothing interesting to say about this that goes beyond Misner's 
original 
observations \cite{Misner} which is why we make the restriction. Since each 
invariant surface contributes a torus to the quotient space we now see that 
our spacetimes can be described as a one parameter family of flat tori; 
spacelike 
tori coming from region I and labelled by ${\tau}$ and Lorentzian tori 
coming from region II and labelled by ${\sigma}$. The Cauchy horizon ${\tau} 
= {\sigma} = 0$ contributes a null torus. 
 
%\begin{figure}
%       \centerline{ \hbox{
%               \epsfig{figure=Torus1.eps,width=10cm}}}
%       \caption{{\small 2+1 dimensional Minkowski space divided into four wedge 
%shaped regions, each of which are foliated by flat surfaces left invariant 
%by ${\Gamma}$. Our covering space consists of regions I and II and our 
%quotient space becomes a one parameter family of tori.}}
%       \label{fig:ett}
%\end{figure}

\begin{figure}[t]
\begin{center}
                \includegraphics[width=8cm]{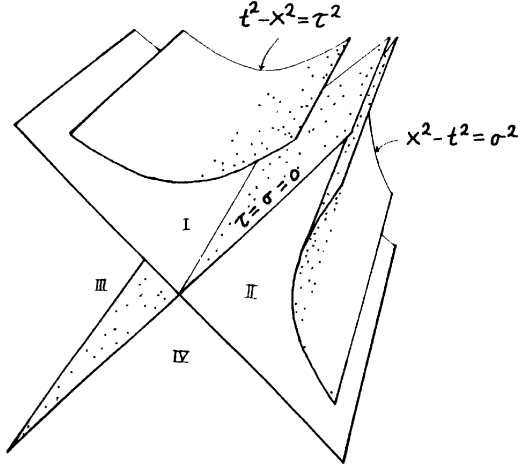}

        \caption{{\small 2+1 dimensional Minkowski space divided into four wedge 
            shaped regions, each of which is foliated by flat surfaces
            left invariant by ${\Gamma}$. Our covering space consists
            of regions I and II and our quotient space becomes a one
            parameter family of tori.}}
        \label{fig:ettett}
\end{center}
\end{figure}

%\vspace{3cm}
%\noindent {\small 2+1 dimensional Minkowski space divided into four wedge 
%shaped regions, each of which are foliated by flat surfaces left invariant 
%by ${\Gamma}$. Our covering space consists of regions I and II and our 
%quotient space becomes a one parameter family of tori.}

\vspace{1cm}

\noindent {\bf 3. A TRIP THROUGH TEICHM\"ULLER SPACE.}

\vspace{5mm}

\noindent Our task now is to describe the one parameter family of flat tori 
that constitutes a spacetime of the kind that we defined in section 2. We use 
the notation that ${\xi}_{\alpha}{\xi}^{\alpha} \equiv ||{\xi}||^2 \equiv \pm 
|{\xi}|^2$, where the sign depends on whether the vector is timelike or 
spacelike and $|{\xi}|$ is non-negative by definition. Let 
us first sketch what goes on in the region without closed timelike curves 
(where a Hamiltonian description is available \cite{Moncrief} 
\cite{Hosoya}). At 
fixed ${\tau}$ the tori are built from parallelograms spanned by the 
generators ${\xi}_1$ and ${\xi}_2$. The angle between them is given by 

\begin{equation} \cos{\theta} = \frac{{\xi}_1\cdot {\xi}_2}{|{\xi}_1||{\xi}_2|} 
= \frac{{\beta}{\delta} + {\alpha}{\gamma}{\tau}^2}{\sqrt{{\beta}^2 + 
{\alpha}^2{\tau}^2}\sqrt{{\delta}^2 + {\gamma}^2{\tau}^2}} \ . \end{equation} 

\noindent Therefore their area is a monotonically increasing function:  

\begin{equation} A = |{\xi}_1||{\xi}_2|\sin{\theta} = ({\alpha}{\delta} - 
{\beta}{\gamma}){\tau} \ . \end{equation}

\noindent (The total area of the torus is the area of a parallelogram times 
a fixed numerical factor that can be chosen at will.) On the other hand 
the shape of the torus is changing in an interesting way. To describe it we 
introduce their Teichm\"uller space: 

\

\noindent \underline{Definition}: Teichm\"uller space is the moduli space 
of marked flat tori. 

\

\noindent "Marked" means that a particular pair of intersecting closed 
geodesics on the torus (namely the one that corresponds 
to our generators ${\xi}_1$ and ${\xi}_2$) has been singled out for special 
attention. The definition applies equally well to Riemannian and Lorentzian 
tori; in the former case it is well known that Teichm\"uller space 
can be regarded as the upper half plane, and that it is naturally equipped 
with the Poincar\'e metric 

\begin{equation} ds^2 = \frac{1}{y^2}(dx^2 + dy^2) \ . \label{6} \end{equation}

\noindent This is hyperbolic space ${\bf H}^2$ and its isometry group is 
$PSL(2, {\bf R})$. We can assign a position in Teichm\"uller space to our 
tori if we first normalize our generators so that ${\xi}_1$ has length one 
and lies along the $x$-axis. Then the tip of ${\xi}_2$ will point at a unique 
point in the upper half plane, namely 

\begin{equation} (x,y) = \frac{|{\xi}_2|}{|{\xi}_1|}(\cos{\theta}, 
\sin{\theta}) = \frac{1}{{\beta}^2 + {\alpha}^2{\tau}^2}({\beta}{\delta} 
+ {\alpha}{\gamma}{\tau}^2, ({\alpha}{\delta} - {\beta}{\gamma}){\tau}) \ . 
\end{equation}

\noindent Note that at this stage we use an auxiliary Euclidean metric 
on the coordinate plane to assign a point to ${\xi}_2$. We now have a 
curve parametrized by ${\tau}$ and it is elementary to show that 
this is a semi-circle meeting the boundary at right angles: 

\begin{equation} \left( x - \frac{{\beta}{\gamma} + {\alpha}{\delta}}
{2{\alpha}{\beta}}\right)^2 + y^2 = \left( \frac{{\alpha}{\delta} - {\beta}
{\gamma}}{2{\alpha}{\beta}}\right)^2 \ . \label{8} \end{equation}

\noindent This is a geodesic with respect to the natural metric. Hence 
the statement that the torus evolves along a geodesic in Teichm\"uller space. 
It should not be forgotten that it also grows in area. A minor calculation 
informs us that if we move a distance $L$ along the geodesic, as measured by 
the Poincar\'e metric, then the area of the torus grows with a factor $e^{L}$. 
Note that this does not depend on the parameters describing the spacetime, nor 
does it depend on where we are on the geodesic.

Now what happens when we pass the Cauchy horizon and enter region II? The first 
observation is that 

\begin{equation} ||{\xi}_1||^2 = {\beta}^2 - {\sigma}^2{\alpha}^2 \ . 
\end{equation}

\noindent Hence (unless ${\xi}_1$ is a pure translation or a pure boost) 
${\xi}_1$ is spacelike in a region where $x^2 - t^2 = {\sigma}^2 < {\beta}^2/
{\alpha}^2$ and it is timelike when $x^2 - t^2 = {\sigma}^2 > {\beta}^2/
{\alpha}^2$. Let us refer to these regions as region IIa and IIb, respectively. 
To avoid misunderstandings, because the group ${\Gamma}$ contains all the 
elements listed in eq. (\ref{xi}) there are closed timelike geodesics 
through every point in region II, although the existence of 
closed null geodesics on the Cauchy horizon depends on whether 
${\delta}/{\beta}$ is rational or not. 

If we now try to mimic the construction of the Teichm\"uller space of Riemannian 
tori we run into a problem with the first step, which was to use a rotation to 
bring the generator ${\xi}_1$ into a standard position. We cannot use Lorentz 
transformations for the same purpose here: The Teichm\"uller space of Lorentzian 
tori splits into two components depending on whether ${\xi}_1$ is spacelike 
or timelike. We therefore use a different approach at first. By definition the 
Teichm\"uller space is the moduli space of marked flat Lorentzian tori. 

\

\noindent \underline{Theorem 1}: The Teichm\"uller space of Lorentzian tori 
has the topology ${\bf R}\times {\bf S}^1$. It is naturally equipped with 
the de Sitter metric.

\

\noindent \underline{Proof}: To each oriented dyad of vectors there corresponds 
a unique flat marked Lorentzian torus. The set of such dyads is isomorphic to 
the group $SL(2, {\bf R})$. If we perform a Lorentz transformation of the dyad 
the torus is unchanged. Taking this into account we find a one-to-one 
correspondence 
between the Teichm\"uller space and the coset space $SL(2, {\bf R})/SO(1,1)$. 
But it is well known that this space has the stated topology. The de Sitter 
metric is natural because it is the maximally symmetric metric, and also 
because it arises if we take the perpendicular distance between the fibers, 
as measured by the standard metric on $SL(2, {\bf R})$. 

\

\noindent Although well known the result is not quite trivial. The coset 
space $SO(2,1)/$ $SO(1,1)$ has the topology of the M\"obius strip, even though 
the group manifolds of $SO(2,1)$ and $SL(2, {\bf R})$ have the same topology. 
Let us give a sketch of the argument:
we may, by analogy with the Euler angle parametrization of ${\bf S}^3$, 
introduce local coordinates $ \th, \ph, \ga$ on $SL(2, {\bf R})$ (aka ${\bf adS}_3$) as

\begin{equation}
\left\{
    \begin{aligned}
      X &= \c \frac{\th}{2} \sh \frac{\ph - \ga}{2} \\
      Y &= \s \frac{\th}{2} \sh \frac{\ph + \ga}{2} \\
      U &= \c \frac{\th}{2} \ch \frac{\ph - \ga}{2} \\
      V &= \s \frac{\th}{2} \ch \frac{\ph + \ga}{2} \\
    \end{aligned}
\right.
\end{equation}

\ni The flat metric

\begin{equation}
ds^2 = dX^2+dY^2-dU^2-dV^2
\end{equation}

on the embedding space induces the metric 

\begin{equation}
ds^2 = \frac{1}{4} \left( -d \th ^2 + d \ph ^2 + d \ga ^2 - 2 d \ph d \ga
\c \th \right)
\label{ads3}
\end{equation}

\ni on $SL(2,{\bf R})$. The coordinate $\ga$ runs along the flow lines of
the Killing field $J_{XU}+J_{YV}$ which generates $SO(1,1)$
transformations and we want to identify points along these lines. The
metric on the resulting space, obtained from the orthogonal distance
between the fibers, may be calculated using the threading approach of
Boersma and Dray\cite{Dray}. By identifying the metric in (\ref{ads3})
with an Ansatz of the form
\begin{equation}
ds^2 = M^2\left( d\ga - M_i dx^i\right) ^2+h_{ij} dx^i dx^j \textrm{,}
\end{equation}
one obtains the metric
\begin{equation}
h = \frac{1}{4} \left( -d\th ^2 + \s^2 \th d\ph ^2 \right)
\end{equation}

\ni for the quotient space $SL(2,{\bf R}) / SO(1,1)$. This is precisely
the metric for (part of) ${\bf adS}_2$ in a reasonably well known
coordinate system; anti-de Sitter space and de Sitter space are
identical in 1+1 dimensions It is also possible to do this calculation in
global coordinates, at the expense of their not being adapted to the
identification Killing field.

As it stands Theorem 1 is not very useful. To see what kind of curve our
tori describe we need to know how to assign a point in Teichm\"uller
space to a given marked torus. This understanding will be provided by
the proof of Theorem 2, which will wind its way to the end of this
section:

\

\noindent \underline{Theorem 2}: The Teichm\"uller space of Lorentzian tori 
has the topology ${\bf R}\times {\bf S}^1$. The one parameter family of 
tori that represents a spacetime (defined in 
section 2) is a timelike geodesic in this space provided that it is equipped 
with the de Sitter metric. 

\

\noindent \underline{Proof}: Our first step is to introduce coordinates 
$(x, t)$. During the construction 
we use the flat Minkowski metric on this coordinate plane. Again we 
normalize the 
vectors so that ${\xi}_1$ points at $(1,0)$. This is always possible provided 
that ${\sigma}^2 < {\beta}^2/{\alpha}^2$. Since we know the scalar product 
of the vectors we find that the tip of ${\xi}_2$ points at the point 

\begin{equation} (x, t) = \frac{1}{{\beta}^2 - {\sigma}^2{\alpha}^2} 
({\beta}{\delta} - {\alpha}{\gamma}{\sigma}^2, - ({\alpha}{\delta} - {\beta}
{\gamma}){\sigma}) \ . \end{equation}

\noindent We have therefore been able to arrange that this component of 
Teichm\"uller space is identical to the lower half plane. It is elementary 
to show that the points on this curve obey

\begin{equation} \left( x - \frac{{\beta}{\gamma} + {\alpha}{\delta}}
{2{\alpha}{\beta}}\right)^2 - t^2 = \left( \frac{{\alpha}{\delta} - {\beta}
{\gamma}}{2{\alpha}{\beta}}\right)^2 \ . \label{14} \end{equation}

\noindent This is a hyperbola with its foci on the $x$-axis and it is a 
geodesic with respect to the metric 

\begin{equation} ds^2 = \frac{1}{t^2}(dx^2 - dt^2) \ . \label{15} \end{equation}

\noindent But this is in fact the de Sitter metric on a coordinate patch 
that covers "one half" of de Sitter space. 

We can now draw a picture of the geodesic in Teichm\"uller space, where 
the Teichm\"uller space of Riemannian tori has been joined to its counterpart 
for Lorentzian tori across their conformal boundaries. Note that in the 
Lorentzian part of the picture the geodesic reaches infinite coordinate values 
at finite parameter values ${\sigma}^2 = {\beta}^2/{\alpha}^2$. This is 
actually a good thing: We know that the coordinates we are using cover only 
a part of Teichm\"uller space. "Infinity" in the picture corresponds to a 
coordinate singularity that is caused by our assumption that ${\xi}_1$ 
is spacelike. 

\begin{figure}[t]
\begin{center}
\includegraphics[width=12cm]{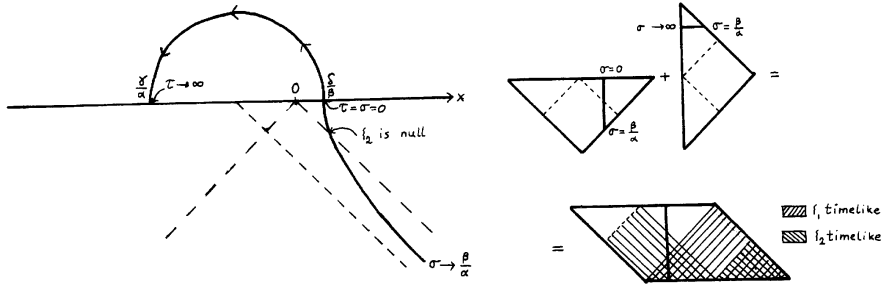}

        \caption{{\small The curve through Teichm\"uller space.
            In the upper half plane the torus is Riemannian. In the
            lower half plane the torus is Lorentzian but the generator
            ${\xi}_1$ is still spacelike. When ${\xi}_1$ is timelike
            we again obtain a half plane. The latter two half planes
            are conveniently depicted with conformal diagrams; adding
            them together so that the curve becomes smooth we obtain
            the conformal diagram of 1+1 dimensional de Sitter
            space.}}
        \label{fig:tva}
\end{center}
\end{figure}

%\vspace{3cm} 
%\noindent {\small The curve through Teichm\"uller space. In the 
%upper half plane the 
%torus is Riemannian. In the lower half plane the torus is Lorentzian but the 
%generator ${\xi}_1$ is still spacelike. When ${\xi}_1$ is timelike we again 
%obtain a half plane. The latter two half planes are conveniently 
%depicted with conformal diagrams; adding them together so that the curve 
%becomes smooth we obtain the conformal diagram of 1+1 dimensional de Sitter 
%space.}

\vspace{5mm}

When ${\xi}_1$ is timelike we again introduce an infinite half plane, this 
time described by the coordinates $t'$ and $x' > 0$, and normalize the vectors 
so that ${\xi}_1$ points at $(t', x') = (1,0)$. We then find that ${\xi}_2$ 
points at 

\begin{equation} (t', x') = \frac{1}{{\sigma}^2{\alpha}^2 - {\beta}^2} 
({\beta}{\delta} - {\alpha}{\gamma}{\sigma}^2, ({\alpha}{\delta} - {\beta}
{\gamma}){\sigma}) \ . \end{equation}

\noindent These points lie on the hyperbola

\begin{equation} {\left( t' + \frac{{\beta}{\gamma} + {\alpha}{\delta}}
{2{\alpha}{\beta}}\right)}^2 - {x'}^2 = {\left( \frac{{\alpha}{\delta} - {\beta}
{\gamma}}{2{\alpha}{\beta}}\right) }^2 \ . \label{17} \end{equation}

\noindent This is a geodesic with respect to the metric 

\begin{equation} ds^2 = \frac{1}{{x'}^2}({dt'}^2 - {dx'}^2) \ . \label{18} 
\end{equation}

\noindent This is again the metric on "one half" of de Sitter space. Since 
we now think of the conformal boundary as being timelike it may be more 
natural to think of it as anti-de Sitter space---but in 1+1 dimensions 
de Sitter space and anti-de Sitter space coincide when we switch the meaning 
of space and time.

It remains to show that the two components of Teichm\"uller space can be 
glued together so that they form a de Sitter space, in such a way that 
the curve becomes a geodesic globally. For this purpose we 
observe that both ${\bf H}^2$ (the Teichm\"uller space of Riemannian tori) 
and 1+1 dimensional de Sitter space can be isometrically mapped into surfaces 
in a 2+1 dimensional Minkowski space with the metric 

\begin{equation} ds^2 = dX^2 + dY^2 - dU^2 \ . \end{equation}

\noindent Explicitly we define an embedding of ${\bf H}^2$ by 

\begin{equation} X = \frac{x}{y} \hspace{8mm} Y + U = \frac{1}{y} \hspace{8mm} 
Y - U = - \frac{x^2 + y^2}{y} \ ; \hspace{5mm} y > 0 \ . \end{equation}

\noindent The surface is the upper sheet of the hyperboloid $X^2 + Y^2 - 
U^2 = - 1$ and the induced metric is the one given in eq. (\ref{6}). The first 
component of the Teichm\"uller space of Lorentzian tori is embedded 
through 

\begin{equation} X = \frac{x}{t} \hspace{8mm} Y + U = \frac{1}{t} \hspace{8mm} 
Y - U = \frac{t^2 - x^2}{t} \ ; \hspace{5mm} t > 0 \ . \end{equation}

\noindent The surface is "one half" of the hyperboloid $X^2 + Y^2 - 
U^2 = 1$ and the induced metric is the one given in eq. (\ref{15}). 
The second component is embedded through 

\begin{equation} X = \frac{t'}{x'} \hspace{8mm} Y + U = - \frac{1}{x'} 
\hspace{8mm} 
Y - U = \frac{{t'}^2 - {x'}^2}{x'} \ ; \hspace{5mm} x' > 0 \ . \end{equation}

\noindent The surface is "the other half" of the hyperboloid $X^2 + Y^2 - 
U^2 = 1$ and the induced metric is the one given in eq. (\ref{18}).

A geodesic in ${\bf H}^2$, and a timelike geodesic in de Sitter space, is 
uniquely defined as the intersection of a hyperboloid with a timelike 
plane through the origin in the embedding space. The curve in Teichm\"uller 
space is given by eqs. (\ref{8}), (\ref{14}) and (\ref{17}). Therefore, to 
show that this curve is globally a timelike geodesic in de Sitter space we 
must find a spacelike vector $k_{\alpha}$ such that eqs. (\ref{14}) and 
(\ref{17}) are equivalent to $k\cdot X = 0$. An elementary calculation 
shows that this is the case for the vector

\begin{eqnarray} (k_X, k_Y, k_U) = ({\alpha}^2{\delta}^2 - {\beta}^2{\gamma}^2, 
\ {\beta}{\delta}({\alpha}^2 + {\gamma}^2) - {\alpha}{\gamma}
({\beta}^2 + {\delta}^2), \ \nonumber \\ 
\ \\ {\alpha}{\gamma}({\beta}^2 - {\delta}^2) + 
{\beta}{\delta}({\gamma}^2 - {\alpha}^2)) \ . \hspace{12mm} \nonumber 
\end{eqnarray}

\noindent Eq. (\ref{8}) is also reproduced. This completes the proof that the 
curve is globally described by a timelike geodesic in de Sitter space. 

\vspace{1cm}

\noindent {\bf 4. THE COGWHEELS OF CHAOS.}

\vspace{5mm} 

\noindent In this section we restrict ourselves to region I (where there 
are no closed timelike curves), so that the evolution can be regarded as 
time evolution in a configuration space in 
the standard sense \cite{Moncrief} \cite{Hosoya}. However, it is a moot 
point whether the configuration space should be taken to be Teichm\"uller 
space or the moduli space of (unmarked) flat tori. The latter space is 
in fact ${\bf H}^2/{\Gamma}_M$, where ${\Gamma}_M$ is the modular group 
$PSL(2, {\bf Z})$ acting on the upper half plane through 

\begin{equation} z \rightarrow z' = \frac{az + b}{cz + d} \ ; \hspace{10mm} 
ad - bc = 1 \label{abcd} \end{equation}

\noindent where $a, b, c$ and $d$ are integers and $z = x + iy$. (To see that 
$z$ and $z'$ actually correspond to the same torus, consider a pair of 
intersecting closed geodesics on the torus and choose them to have the shortest 
circumference possible. The conformal structure can be characterized by 
the angle and relative lengths of this pair. A little experimentation 
shows that these are unaffected by a modular transformation.) The quotient 
space is the famous modular surface, usually described as the fundamental 
region of the group which is bounded 
by $r^2 \equiv x^2 + y^2 = 1$ and $x = \pm 1/2$. It is depicted in fig. 3. 
Its area is finite and it is a smooth manifold 
except for two conical singularities occurring at the fixed points of the 
transformations $S$ and $ST$, where $S$ and $T$ are the transformations

\begin{equation} Sz = - \frac{1}{z} \hspace{12mm} Tz = z + 1 \ . \end{equation}

\noindent $S$ and $T$ generate the group and obey two relations, viz. $S^2 = 1$ 
and $(ST)^3 = 1$. Note that the transformation $S$ acts by switching the 
elements in the oriented dyad that defines the torus. 

\begin{figure}
\begin{center}
\includegraphics[width=11cm]{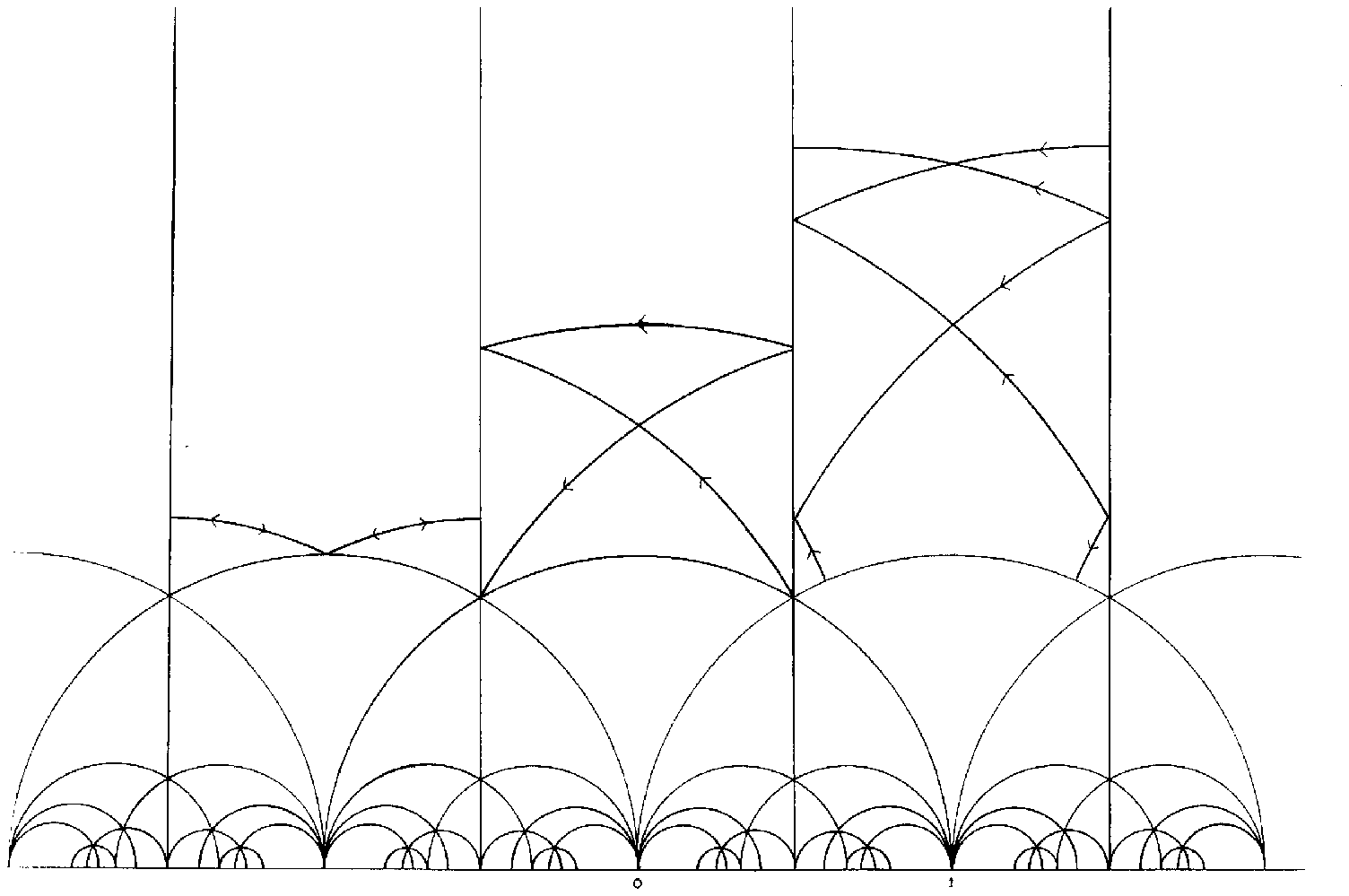}
        \caption{{\small The modular surface is the fundamental 
            region of the modular group, with sides appropriately
            identified. The picture shows how the upper half plane is
            tesselated by copies of the fundamental region. In three
            of the copies we have drawn examples of closed geodesics
            ($N = 3$ $x_+ = [\dot{1}]$, $N = 4$ $x_+ = [\dot{1},
            \dot{2}]$ and $N = 5$ $x_+ = [\dot{1}, \dot{3}]$ in the
            notation introduced below).}}
        \label{fig:tre}
\end{center}
\end{figure}

%\vspace{3cm}
%\noindent {\small }
%\vspace{5mm}

The question whether the configuration space is ${\bf H}^2$ or ${\bf H}^2/
{\Gamma}_M$ matters for the properties of the model but it is not a 
question of right or wrong, since we do not intend to compare the model to 
experiment anyway. Technically the modular group does not belong to the 
connected component of the gauge group so that both options are open as far 
as consistency is concerned. For thoughtful comments on this issue we refer to 
papers by Peld\'an \cite{Peldan} and Matschull \cite{Matschull}; here we 
choose the second option because it is an interesting one. 

As shown by Artin \cite{Artin} and Hedlund \cite{Hedlund} the geodesic flow 
on the modular surface is ergodic (indeed they showed this at a time when 
the proper definition of an ergodic system was yet to be found---with 
today's definition we can say that the flow has the Bernoulli property, 
which is the strongest ergodic property around). From this point of view it 
has been much studied; Series has written a nice review with some entries 
to the technical literature \cite{Series}. Here we focus on one aspect of 
this flow, namely its closed orbits. We take the point of view that one can 
define "chaos" in a dynamical system by the requirement that the number of its 
unstable closed orbits rises exponentially as a function of length. This is not 
at all unreasonable; in fact this is the feature of chaotic systems that 
survives the transition to quantum theory (via the Gutzwiller trace 
formula, which connects 
the asymptotic properties of the spectrum of closed geodesics to the spectrum 
of the Laplacian). Since it is a simple matter of counting it is also a feature 
that survives the transition to diffeomorphism invariant systems---unlike 
Lyapounov exponents and the like that can be reparametrized away. To avoid 
confusion, note that---because the area of our tori is growing---a closed 
geodesic in moduli space actually corresponds to a self-similar rather than 
a periodic spacetime.

The closed geodesics on the modular surface arise because any hyperbolic 
M\"obius transformation---corresponding to an $SL(2, {\bf R})$ matrix 
whose trace has an absolute value larger than two---has a unique geodesic 
flowline connecting its pair of fixed points on the real axis. If this 
M\"obius transformation is a modular transformation as well there are 
points on this geodesic that will be identified with each other, and 
a closed geodesic results. The distance $L$ between 
a pair of neighbouring identified points is easily computed. It is given by 

\begin{equation} 2\cosh{\frac{L}{2}} = N \ , \label{26} \end{equation}

\noindent where $N = |\mbox{Tr}g|$ and $g$ is the matrix corresponding to the 
modular transformation, so that $N = a + d$ if the transformation is written as 
in eq. (\ref{abcd}). Note that $N$ can be used to label the conjugacy classes 
of $SL(2, {\bf R})$. This therefore is the length spectrum of the closed 
geodesics. 

It takes more effort to understand how many closed geodesics there are. In group 
theoretical terms this is the problem to enumerate the conjugacy classes of 
$PSL(2, {\bf Z})$. There are only two conjugacy classes of elliptic elements, 
corresponding to the two fixed points on the boundary of the fundamental region. 
The number of conjugacy classes of hyperbolic elements on the other hand 
is a rapidly growing function of $N$. It is in fact known (see for instance 
ref. \cite{Bogomolny}) that when $L$ is large the number $n$ of closed 
geodesics with length $l$ not exceeding $L$ grows like 

\begin{equation}  n(l \leq L) \sim \frac{e^L}{L} \ . \label{19} \end{equation}

\noindent This settles it: The system is chaotic. It is however an instructive 
exercise to compute the number of closed geodesics "from below" with pedestrian 
methods, and this we will now proceed to do. 

A geodesic in the upper half plane can be conveniently characterized by two real 
numbers, its starting point $x_+$ and its end point $x_-$ on the real 
axis. To each geodesic we can associate a hyperbolic M\"obius transformation 
whose fixed points are these two points. The geodesic projects to a closed 
geodesic on the modular surface if and only if this M\"obius 
transformation belongs to the modular group, and there will be a unique 
such M\"obius 
transformation of smallest trace associated to the closed geodesic (if $x = 
gx$ then $x = g^nx$; if $n > 1$ the trace of $g^n$ is greater than 
the trace of $g$ and the corresponding geodesic is traversed several 
times---here 
we count only "primitive" closed geodesics). In equations then 

\begin{equation} x_{\pm } = \frac{ax_{\pm } + b}{cx_{\pm } + d} \ . 
\end{equation}

\noindent It follows that $x_{\pm }$ is a quadratic surd, that is to say a 
solution to a quadratic algebraic equation with integer coefficients whose 
discriminant is not a perfect square. The two solutions to this equation are 

\begin{equation} x_{\pm } = \frac{1}{2c}(a - d \pm \sqrt{N^2 - 4}) \ , 
\label{29} \end{equation}

\noindent where $N = a + d$ and we made use of the condition $ad - bc = 1$. 
Note that the discriminant $D = N^2 - 4 = 4\sinh^2{\frac{L}{2}}$ according 
to eq. (\ref{26}). Since the surds occur in pairs the closed geodesics 
can in fact be labelled by just one real number, say its "source" $x_+$.

Next we introduce continued fractions \cite{Rockett}. A real number can be 
uniquely expressed in the form 

\begin{equation} x = a_0 + \frac{1}{a_1 + \frac{1}{a_2 + \ ...}} 
\equiv [a_0, a_1, a_2, \ ... ] \ , \end{equation}

\noindent where all the partial quotients $a_i$ are integers and all except 
possibly $a_0$ are positive. It is known that $x$ is rational if and only if 
its continued fraction expansion is finite (i.e. the number of its partial 
quotients is finite), and it is a quadratic surd if and only if its continued 
fraction expansion eventually repeats, in which case it is called periodic. 
The beginning and end of the period is then marked with overdots, so that 
a quadratic surd of period length $k$ is of the form $x = [a_0; a_1, \ ... \ 
a_{n-1}, \dot{a}_{n}, \ ... \ , \dot{a}_{n+k-1}]$. This nice characterization 
of quadratic surds is interesting to us.

One piece of the technology of continued fractions should be mentioned, which 
is that they give rise to a sequence of approximations of $x$ by rational 
numbers: 

\begin{equation} [a_0] = \frac{p_0}{q_0} \hspace{1cm} [a_0,a_1] = 
\frac{p_1}{q_1} 
\hspace{1cm} [a_0, a_1, a_2] = \frac{p_2}{q_2} \end{equation}

\noindent and so on. Here $p_n$ and $q_n$ are polynomials in the partial 
quotients 
and by induction one can show that 

\begin{equation} p_n = a_np_{n-1} + p_{n-2} \hspace{1cm} q_n = a_nq_n + 
q_{n-1} \ . \end{equation}

\noindent Note that $p_n$ and $q_n$ are monotonically increasing functions of 
$n$. 

We want to count equivalence classes of geodesics under the modular 
group and therefore we will try to fix one member of each equivalence class. 
Now the modular group acts on a continued fraction in the following way:

\begin{eqnarray} x = [a_0, a_1, a_2, a_3, \ ... \ ] \rightarrow 
ST^{-a_0}x = - [a_1, a_2, a_3, \ ... \ ] \rightarrow\nonumber \\ \ \\ 
\rightarrow ST^{a_1}ST^{- a_0}x = [a_2, a_3, \ ... \ ] \ . \nonumber 
\end{eqnarray}

\noindent It follows that we can remove the partial quotients in pairs. In 
particular it follows that we can choose $x_+$ to be a purely periodic continued 
fraction since we can always remove the initial sequence. Hence without loss of 
generality

\begin{equation} x_+ = [\dot{a}_0, \ ... \ , \dot{a}_{k-1}] \ . \end{equation}

\noindent If the period length $k$ is even then $x_+$ is a fixed point of the 
group element 

\begin{equation} g = ST^{a_{k-1}} \ ... \ ST^{a_1}ST^{-a_0} \ . \end{equation}

\noindent In terms of the polynomials introduced above it can be shown that 

\begin{equation} x_+ = gx_+ = \frac{q_{k-2}x_+ - p_{k-2}}{- q_{k-1}x_+ + 
p_{k-1}} \hspace{3mm} \Rightarrow \hspace{3mm} N = |\mbox{Tr}g| = p_{k-1} + 
q_{k-2} \ . \label{36} \end{equation}

\noindent This is a useful fact since it means that $N$ is a monotonically 
increasing function of the partial quotients. It also means that $N$ will 
grow when the length of the period in the continued fraction grows, other 
things being equal. 

The fixed point $x_+$ is in fact the source of the geodesic associated with 
$g$. This is so because $g$ removes one period from the continued fraction, 
so that when $g$ acts on an approximation to $x_+$ that is a rational 
number whose continued fraction expansion consists of a finite number 
of periods then $g$ 
moves that rational number away from $x_+$. If the period length is odd 
then that $g$ which leaves it fixed and has the smallest value of $N$ is 

\begin{equation} g =  ST^{a_{k-1}}\cdot \ ... \ \cdot ST^{a_0}ST^{- a_{k-1}} 
\cdot \ ... \ \cdot ST^{- a_0} \ . \end{equation}

\noindent It is convenient to regard continued fractions of odd period 
lengths as 
having even periods of twice the original length. According to a theorem of 
Galois' the corresponding sink (the other root of the quadratic equation) 
now obeys 

\begin{equation} Sx_- = - \frac{1}{x_-} = [\dot{a}_{k-1}, \ ... \ , 
\dot{a}_0] \ . \end{equation}

\noindent It is easy to show this since $x_-$ is the source of the group element 
$g^{-1}$. The source and sink are now given in reduced form; this means that 
$x_+ > 1$ and $-1 < x_- < 0$. 

Two geodesics in reduced form will give rise to the same closed geodesic on the 
modular surface if one can be obtained from the other by cyclic permutations 
of the pairs in the continued fraction expansion of their sources. This 
remaining ambiguity is easy to take care of, so that we can now make a 
list of all closed geodesics corresponding to continued fraction 
expansions of a given period length. 
Moreover we know from eq. (\ref{36}) that the length of the geodesic is a 
monotonically increasing function of the partial quotients, so it is 
straightforward to compute the number of primitive closed geodesics of a length 
not exceeding some chosen reasonable number. The result of such a calculation is 
given in fig. 4. Continuing this exercise on a computer one can see how eq. 
(\ref{19}) emerges. (Curiously we were unable to find this calculation in 
the accessible literature, although it has been done before \cite{Diploma}.)

\vspace{3cm}
\begin{figure}
\begin{center}
  \includegraphics[width=8cm]{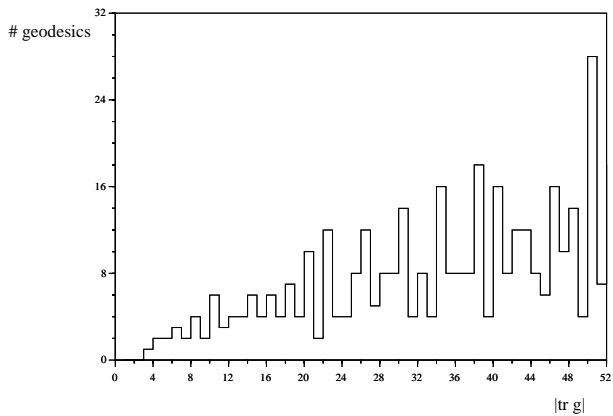}
        \caption{{\small Degeneracies of the length spectrum: The first 52 levels.}}

\end{center}
\end{figure}

\vspace{5mm}

The conclusion is that 2+1 gravity on the torus is a chaotic system according 
to the definition that we have adopted. Unlike the case of Bianchi models no 
approximation was involved. It may be felt that this chaos was introduced by 
sleight-of-hand since the system was in fact integrable before the modular 
group was declared to generate gauge symmetries. Indeed we are 
dealing with chaos of a very special kind, called "arithmetical chaos". Although 
the system is chaotic in the sense that the number of closed orbits not 
exceeding a 
given length grows exponentially, it is also very special because there are 
huge degeneracies in the length spectrum (caused by the fact that the number 
of possible lengths grows much more slowly). Closer investigation reveals 
\cite{Bogomolny} that in such situations the level statistics of the Laplace 
operator shows some features that resemble integrable systems much more than 
they resemble a generic chaotic system (in particular the level repulsion that 
is typical of the latter is missing here) so the feeling is justified to some 
extent.

\vspace{1cm}

\noindent {\bf 5. A LOOSE END.}

\vspace{5mm}

\noindent In the previous section we occupied ourselves with the action of 
the modular group on the Teichm\"uller space of Riemannian tori; the quotient 
space---the moduli space of Riemannian tori---is almost a smooth manifold 
since the modular group has only two elliptic conjugacy classes, and only 
the elliptic members of the modular group have fixed points in ${\bf H}^2$. 
The situation is dramatically different for the action of the modular group 
on the Teichm\"uller space of Lorentzian tori: Here every hyperbolic element 
of the modular group has fixed points inside the space, and we have already 
seen that there is an infinite number of inequivalent elements of this type. 

The modular group is a subgroup of $PSL(2, {\bf R}) = SO(2,1)$ and this is 
the isometry group of ${\bf H}^2$ and 1+1 dimensional de Sitter space alike. 
The action of the generators of the modular group is as follows. The generator 
$T$ gives rise to a "null rotation" generated by a Killing vector that becomes 
null along the coordinate singularity that separates the two parts of 
de Sitter space in the description we gave above; its fixed points lie 
on the conformal boundary. In the half plane coordinates it is simply 
a translation in the $x$-direction. The generator $S$ is a spatial rotation of 
de Sitter space; it has no fixed points and cannot be described in a single 
coordinate patch of the type used above. If we think of $S$ as effecting an 
interchange of the basis elements in the dyad that defines the torus we 
see that this must be so whenever one of the elements is spacelike and the 
other timelike---the generator $S$ will then transform a point representing 
a spacelike ${\xi}_1$ (say) into a point in the other coordinate patch where 
${\xi}_1$ is timelike. Fig. 5 should be enough to make this clear. 

\begin{figure}
\begin{center}
\includegraphics[width=9cm,clip=true]{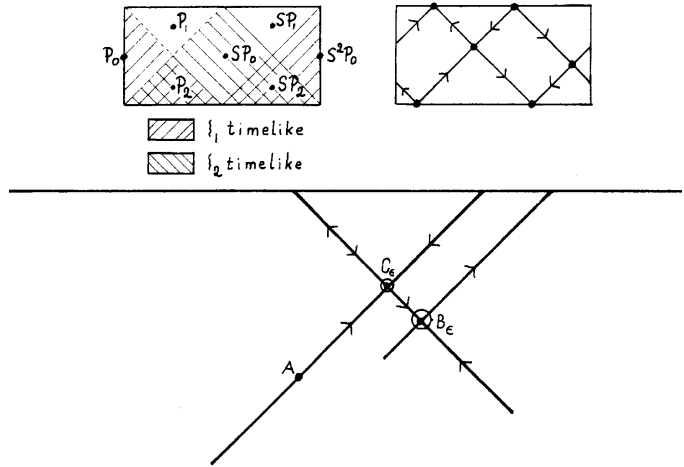}
        \caption{{\small The action of $S$ on 1+1 dimensional de Sitter space; 
            also the null flow lines and the fixed points of a
            hyperbolic transformation; and a sketch of the proof
            (involving two different hyperbolic transformations) that
            the action of the modular group is ergodic.}}
        \label{fig:fem}
\end{center}
\end{figure}
%\vspace{3cm}
%\noindent {\small The action of $S$ on 1+1 dimensional de Sitter space; 
%also the null flow lines of a hyperbolic transformation; and a sketch of 
%the proof (involving two different hyperbolic transformations) that the 
%action of the modular group is ergodic.}
%\vspace{5mm}

Each hyperbolic element of the modular group has two fixed points inside 
de Sitter space, and two fixed points on each component of the conformal 
boundary separated from the fixed points in the interior by null lines 
that are left invariant by the transformation. The fixed points on the 
boundary are conjugate pairs of quadratic surds and conversely. This 
makes it easy to prove the next theorem: 

\

\noindent \underline{Theorem 3}: The action of the modular group on the 
Teichm\"uller space of Lorentzian tori is ergodic, in the sense that 
an arbitrary point can be transformed into an arbitrary coordinate 
neighbourhood of any other point. 

\

\noindent \underline{Proof}: We must show that there is a modular 
transformation taking an arbitrary point $A$ into a given neighbourhood 
$B_{\epsilon}$ of another arbitrary point $B$. All neighbourhoods 
are regarded as coordinate neighbourhoods and we assume that the pair 
of points lies within some half plane coordinate patch. (There are 
exceptional pairs for which this fails, but they can easily be 
treated with an extension of the argument and will be ignored.) 

We need to know that in any neighbourhood of any point there is a 
hyperbolic modular transformation with a fixed point in that neighbourhood. 
This will be so if, given any two points $x_{\pm }$ on the conformal boundary, 
we can find a conjugate pair of quadratic surds with one member 
arbitrarily close to each. But this is easy using the 
technology of the previous section. First modular transformations are 
used to show that it is enough to consider the case $x_+ > 1$, 
$- 1 < x_- < 0$. Then one approximates $x_+$ and $ - 1/x_-$ 
with continued fractions to the desired accurary. The sequence of 
integers that gives the continued fraction approximating $- 1/x_-$ is 
then reversed and added to the sequence that approximates $x_+$, 
and the resulting sequence is taken to be the period of a purely 
periodic continued fraction. Galois' theorem shows that we now have 
an approximation of $x_+$ whose conjugate surd approximates $x_-$. At 
the end we choose $x_{\pm}$ to be null separated from the given point. 
They intersect at a fixed point, and we are done. 

With this understanding, draw null lines through $A$ and $B$ meeting 
each other at the point 
$C$. Choose a suitable neighbourhood $C_{\epsilon}$ of $C$ and a 
hyperbolic modular transformation with a fixed point in $C_{\epsilon}$. 
Use this transformation to move the point $A$ into $C_{\epsilon}$. 
Then choose a hyperbolic modular transformation with a fixed point 
in $B_{\epsilon}$ and adjust the size of $C_{\epsilon}$ so that the 
second transformation moves $C_{\epsilon}$ into $B_{\epsilon}$. 

\

\noindent Except for a speculative remark in the conclusions we have 
nothing to say about what this means. 

\vspace{1cm}

\noindent {\bf 6. OTHER SPACETIMES.}

\vspace{5mm}

\noindent The final issue is to what extent the results described above 
are peculiar to flat spacetimes and to the genus one case. We confine our 
remarks to region I, where there are no closed timelike 
curves and the tori are spacelike. Consider first locally de Sitter spacetimes. 
2+1 dimensional de Sitter spacetime can be described as the hypersurface 

\begin{equation} X^2 + Y^2 + Z^2 - U^2 = 1 \end{equation}

\noindent embedded in a four dimensional Minkowski space (with $U$ as its 
time coordinate). Alternatively, it is the maximally symmetric vacuum solution 
to Einstein's equations with a positive cosmological constant ${\lambda}$. 
Again we choose two commuting and linearly independent Killing vectors 

\begin{equation} {\xi}_1 = {\alpha}J_{ZU} + {\beta}J_{XY} \hspace{8mm} 
{\xi}_2 = {\gamma}J_{ZU} + {\delta}J_{XY} \ . \end{equation}

\noindent They leave invariant the flat surfaces 

\begin{equation} U^2 - Z^2 = \sinh^2{\tau} \ , \end{equation}

\noindent whose mean curvature is $K = 4\cosh{\tau}$. Following the same steps 
as above we find that the invariant flat surfaces are turned into tori and 
that the evolution of the shape of these tori is given by a geodesic in 
Teichm\"uller space, with the interesting difference \cite{Soda} that the 
evolution slows down and tends to a definite point in Teichm\"uller 
space as the parameter ${\tau}$ goes to infinity (while the area continues 
to grow). Explicitly

\begin{equation} (x, y) = \frac{1}{{\alpha}^2\tanh^2{\tau} + 
{\beta}^2}({\alpha}{\gamma}\tanh^2{\tau} + {\beta}{\delta}, 
({\alpha}{\delta} - {\beta}{\gamma})\tanh{\tau})  \end{equation} 

\begin{equation} A = ({\alpha}{\delta} - {\beta}{\gamma})
\sinh{\tau}\cosh{\tau} \ . \end{equation}

\noindent The evolution stops because $\tanh{\tau} \rightarrow 1$ as 
${\tau} \rightarrow \infty $. Note that this time the change of area as 
we move a distance $L$ along the geodesic does depend on where we are 
on the geodesic. A subtlety should be mentioned also, namely that the 
universal covering space of the quotient spaces considered here is not, 
in general, de Sitter space itself but a "larger" incomplete spacetime 
of constant curvature \cite{Mess} \cite{Ezawa}; for the best explanation 
that we have to offer see ref. \cite{Holst}. 

Why does the evolution stop in the interior of Teichm\"uller space? The answer 
is in fact obvious: In the de Sitter case future infinity \scri \ is a spacelike 
surface transformed into itself by ${\Gamma}$. When we take the 
quotient we obtain an 
"asymptotic torus" with a definite conformal structure, and this is the endpoint 
of the geodesic in Teichm\"uller space. The area of this torus is not defined 
since \scri \ is equipped with a conformal structure only. At this point the 
reader may object that \scri \ is a sphere and that a discrete group like 
our ${\Gamma}$ cannot act properly discontinuously on a sphere. This is true 
but irrelevant; in fact the covering space that we are using is not quite de 
Sitter space but an incomplete spacetime obtained by removing two timelike lines 
from de Sitter space, and afterwards going to the universal covering space. 
This means that \scri \ is really a twice punctured sphere that has been 
"unrolled" to form a plane. This is explained in fig. 7 in ref. \cite{Holst}, 
where it can be seen that the invariant flat spacelike surfaces that were 
defined in the previous section do not encounter the timelike lines that 
were removed (except on \scri \ itself).

For the genus one case then we find that the chaotic behaviour in the 
moduli space of tori is somehow "washed away" by the cosmological constant. 
It should however be noted that the flat torus universe is quite special in 
this regard. We can obtain locally flat spacetimes foliated by Riemann 
surfaces of higher genus by choosing ${\Gamma}$ to be a discrete group---but 
this time not a free group---generated by non-commuting elements that in 
general are combinations of boosts and translations. These spacetimes are 
conformally static when ${\Gamma}$ consists of pure boosts. As time passes 
the boost parts will dominate the translations and the solution will tend to a 
conformally static solution, that is to a definite point inside Teichm\"uller 
space. (This has been demonstrated with full rigour \cite{Larsa}.) 
For the genus one case the evolution never stops for essentially 
the same reason; it is still true that eventually the boost part of the 
generators will dominate but now this means that the shape of the torus 
degenerates so that we approach the boundary of Teichm\"uller space. 

\vspace{1cm}

\noindent {\bf 6. CONCLUSIONS.}

\vspace{5mm}

\noindent The main new results of this paper are the explicit description 
of the moduli space of Lorentzian tori as the union of two half planes 
constituting a 1+1 dimensional de Sitter space, and the demonstration 
that the description of the 2+1 dimensional locally flat torus universe 
as a geodesic in Teichm\"uller space is valid on both sides of the Cauchy 
horizon. We also emphasized the analogy between these 2+1 dimensional 
spacetimes on the one hand, and mixmaster cosmology on the other. The 
difference between them is that the BKL approximation is exact in the 
former case. This is interesting because it shows that chaotic behaviour 
in general relativity should not in general be blamed on strong 
gravitational fields. 

There are some open ends. We did not describe the extension to a geodesically 
complete spacetime, but this was mainly because it appears clear that 
this would give nothing new (compared to Misner's original work 
\cite{Misner}). A more interesting open end is that the analogy to 
Bianchi IX cosmology holds only in the region where there are no 
closed timelike curves and the configuration space can be taken 
to be the moduli space of flat Riemannian tori, which is almost a 
smooth manifold. In the region with closed timelike curves we have 
to deal with the moduli space of Lorentzian tori, which is defined 
as the quotient of 1+1 dimensional de Sitter space by the modular 
group. But---as we demonstrated---the action of the modular group 
is now ergodic, so that the resulting quotient space is not easily 
described even as a set. It is our understanding that the desire 
to describe sets of this type is one of the main motivations behind 
non-commutative geometry \cite{Connes}. It would be marvellous if 
one could follow this lead in such a way that an analogy with the 
singularity in 3+1 dimensional cosmologies could be drawn. 

\vspace{1cm}

\noindent \underline{Acknowledgements}:

\

\noindent We learned much from many people while thinking about these 
issues. Notably from Hans-J\"urgen Matschull, S\"oren Holst, Torsten Ekedahl, 
Lars Andersson and Stefan \AA minneborg.


\newpage

\begin{thebibliography}{99}

\bibitem{Carlip} S. Carlip: Quantum Gravity in 2+1 Dimensions, Cambridge 
U.P. 1998.

\bibitem{Moncrief} V. Moncrief, J. Math. Phys. \underline{30} (1989) 2907.

\bibitem{Hosoya} A. Hosoya and K. Nakao, Class. Quant. Grav. \underline{7} 
(1990) 163.

\bibitem{Belinski} V.A. Belinskii, I.M. Khalatnikov and E.M. Lifshitz, 
Adv. in  Physics \underline{19} (1970) 525.
 
\bibitem{Charles} C.W. Misner, in J. Klauder (ed.): Magic without Magic, 
Freeman, San Fransisco 1972.

\bibitem{Hobill} D. Hobill, A. Burd and A.A. Coley (eds.): Deterministic 
Chaos in General Relativity, Plenum Press, N.Y. 1994. 

%\bibitem{Barrow} J. Barrow, Phys. Rep. \underline{85} (1982) 1.

%\bibitem{Rugh} S. Rugh

\bibitem{Ringstrom} H. Ringstr\"om, The Bianchi IX attractor, 
gr-qc/0006035, 2000.

\bibitem{Moore} G. Moore, Finite in All Directions, hep-th/9305139, 
1993.

\bibitem{Mess} G. Mess, Lorentz spacetimes of constant curvature, IHES 
preprint M/90/28, 1990.

\bibitem{Louko} J. Louko and D. Marolf, Class. Quant. Grav. \underline{11} 
(1994) 311.

\bibitem{Yurtsever} U. Yurtsever, J. Math. Phys. \underline{31} (1990) 
311.

\bibitem{Artin} E. Artin, Collected Papers p. 499, Addison-Wesley 1965.

\bibitem{Series} C. Series, Math. Intelligencer \underline{4} (1985) 24.

\bibitem{Bogomolny} E.B. Bogomolny, B. Georgeot, M.-J. Giannoni and 
C. Schmit, Phys. Rep. \underline{291} (1997) 219.

\bibitem{Soda} Y. Fujiwara and J. Soda, Prog. Theor. Phys. \underline{83} 
(1990) 733.

\bibitem{Misner} C.W. Misner, in J. Ehlers (ed.): Relativity Theory and 
Astrophysics, Amer. Math. Soc., Providence, R.I. 1967.

\bibitem{Dray} S. Boersma and T. Dray, Gen. Rel. Grav. \underline{27} 
(1995) 319.

\bibitem{Peldan} P. Peld\'an, Phys. Rev. \underline{D53} (1996) 3147.

\bibitem{Matschull} H.-J. Matschull, Class. Quant. Grav. \underline{16} 
(1999) 2599.

\bibitem{Hedlund} G. A. Hedlund, Am. J. Math. \underline{57} (1935) 668.

\bibitem{Rockett} A. M. Rockett and P. Sz\"usz: Continued Fractions, 
World Scientific 1992

\bibitem{Diploma} D. Schleicher, Diploma Thesis, University of Hamburg 
1991.

\bibitem{Ezawa} K. Ezawa, Phys. Rev. \underline{D49} (1994) 5211.

\bibitem{Holst} I. Bengtsson and S. Holst, Class. Quant. Grav. \underline{16} 
(1999) 3735.

\bibitem{Larsa} L. Andersson, V. Moncrief and A.J. Tromba, J. Geom. 
Phys. \underline{23} (1997) 191.

\bibitem{Connes} A. Connes, Noncommutative Geometry, Academic Press 1994.

\end{thebibliography}
\end{document}